\documentclass{emulateapj}
\usepackage{graphicx}
\usepackage{psfig}
\usepackage{natbib}

\newcommand{\RA}[3]{{#1}^{{\rm h}}{#2}^{{\rm m}}{#3}^{{\rm s}}}
\newcommand{\Dec}[3]{{#1}^{\circ}{#2}'{#3}''}
\newcommand{\E}[1]{\times 10^{#1}}
\newcommand{\twCO}{$^{12}$CO}  \newcommand{\thCO}{$^{13}$CO}

\newcommand{\HII}{\mbox{H\,\textsc{ii}}}

    \newcommand{\Msun}{M_{\odot}}

     \newcommand{\du}{d_{4.4}}
\newcommand{\VLSR}{V_{\rm LSR}}

\begin{document}

\title{
INTERACTION BETWEEN SUPERNOVA REMNANT G22.7$-$0.2 AND
THE AMBIENT MOLECULAR CLOUDs
}

\shorttitle{Shock--MC Interaction in SNR G22.7$-$0.2}

\author{
Yang Su\altaffilmark{1,2}, Ji Yang\altaffilmark{1,2}, 
Xin Zhou\altaffilmark{1,2}, Ping Zhou\altaffilmark{3}, 
and Yang Chen\altaffilmark{3}
       }

\affil{
$^1$ Purple Mountain Observatory, Chinese Academy of
Sciences, Nanjing 210008, China \\
$^2$ Key Laboratory of Radio Astronomy, Chinese Academy of
Sciences, Nanjing 210008, China \\
$^3$ Department of Astronomy, Nanjing University, 163 Xianlin Avenue,
Nanjing 210023, China \\
      }

\begin{abstract}
We have carried out \twCO\ ($J$=1--0 and 2--1), \thCO\ ($J$=1--0), and 
C$^{18}$O ($J$=1--0) observations in the direction of the 
supernova remnant (SNR) G22.7$-$0.2. A filamentary molecular gas 
structure, which is likely part of a larger molecular complex 
with $\VLSR\sim$75--79~km~s$^{-1}$, is detected and is found to surround 
the southern boundary of the remnant. In particular, the high-velocity wing 
(77--110~km~s$^{-1}$) in the \twCO\ ($J$=1--0 and $J$=2--1) emission shows 
convincing evidence of the interaction between 
SNR G22.7$-$0.2 and the 75--79~km~s$^{-1}$ molecular clouds (MCs). 
Spectra with redshifted profiles, a signature of shocked molecular gas,
are seen in the southeastern boundary of the remnant.
The association
between the remnant and the 77~km~s$^{-1}$ MCs places the remnant 
at the near distance of 4.4$\pm$0.4~kpc,
which agrees with a location on the Scutum--Crux arm.
We suggest that SNR G22.7$-$0.2, SNR W41, and \HII\ region G022.760-0.485
are at the same distance and are associated with GMC G23.0$-$0.4.

\end{abstract}


\section{INTRODUCTION}
Massive stars form from molecular clouds (MCs) and they interact 
with their ambient interstellar medium (ISM). Mass and energy deposition
from the massive stars into the surrounding ISM occurs via strong
ultraviolet radiation, powerful stellar winds, and finally,
violent supernova explosions.
When these blast waves hit the 
nearby MCs, the shocks will compress, heat, accelerate, and even dissociate
the molecular gas, which leads to a wide variety of observable effects.
The interplay between the shock and the molecular gas may trigger 
star formation in the nearby giant molecular cloud (GMC).
A shock interaction with MCs can also generate $\gamma$-rays as a
result of a neutral pion decay after a $p$-$p$ collision (hadronic interaction),
which may exhibit observable very high energy (VHE) emissions.
In all of the study field, the molecular emission is a
useful tool to investigate the nature of shock--MC 
interactions \cite[see a recently view in][]{2014IAUS..296..170C}.

Supernova remnant (SNR) G22.7$-$0.2, with a diameter of 26$'$, is 
listed in the catalog of \cite{1991PASP..103..209G}. 
The remnant is located in a complex
field with multiple \HII\ regions and an adjacent SNR G23.3$-$0.3 (W41)
\cite[see Figures~11 and 12 in][]{2010ApJ...708.1241M}
that has been studied less.
SNR G22.7$-$0.2 shows a faint shell structure in the radio emission and
it displays a concave structure on the southern part of the shell.
There are numerous 
mid-infrared filaments evident at 5.8--8$\mu$m in coincidence with the 
radio shell of the remnant, which was suspected as a possible region
of SNR--MC interaction \citep{2006AJ....131.1479R}. 
\cite{2006AJ....131.2525H} identified the G22.7$-$0.2's southern object 
G22.7583$-$0.4917 as a SNR candidate with 5\farcm0 diameter, while 
\cite{2006A&A...453.1003T} suggested that G22.76$-$0.49 is probably an 
\HII\ region. Including the \HII\ regions or SNR candidates
in the field of view (FOV) of SNR G22.7$-$0.2 \citep{2006AJ....131.2525H,
2014ApJS..212....1A}, a VHE source, HESS J1832--093, is lying on the western 
edge of SNR G22.7$-$0.2, but its origin remains undetermined
\citep{2013arXiv1308.0475L}.

The study of SNR G22.7$-$0.2 is motivated by the possible relationship 
between the multiple objects and the remnant in the complex region. 
We performed new millimeter CO ($J$=1--0) observations (covering 
about 1 deg$^2$) toward the remnant to investigate the overall 
MC distribution surrounding the complex region. 
Very wide 77--110~km~s$^{-1}$ line broadenings are seen in the spectra
of \twCO\ ($J$=1--0 and $J$=2--1) in the southeastern boundary of the 
remnant; this represents evidence for shock--MC interactions.
We suggest that the $\VLSR\sim$75--79~km~s$^{-1}$ MCs are physically
associated with the remnant because of the spatial and kinematic features.
We also explore the relationship between the 
SNR G22.7$-$0.2 and the overlapping \HII\ regions.

The paper is structured as follows. In Section 2, we show the CO
observations and the data reduction. In Sections 3 and 4, we describe
the main results and the physical discussion, respectively. A brief summary
is given in Section 5.

\section{OBSERVATIONS AND DATA REDUCTION}
The observations toward SNR G22.7$-$0.2 were made simultaneously in the
\twCO~($J$=1--0) line (at 115.271~GHz), the \thCO~($J$=1--0) line
(110.201~GHz), and the C$^{18}$O~($J$=1--0) line (109.782~GHz) during 2013
March and April using the 13.7 m millimeter-wavelength telescope
of the Purple Mountain Observatory at Delingha in China.
It is a part of the Milky Way Imaging Scroll
Painting (MWISP) project for investigating the 
nature of the molecular gas along the northern Galactic Plane.
We used a new 3$\times$3 pixel Superconducting Spectroscopic Array 
Receiver as the front end, which was made with 
Superconductor--Insulator--Superconductor (SIS) mixers using the 
sideband separating scheme \citep{shan, 2011AcASn..52..152Z}. 
An instantaneous bandwidth of 1~GHz was used as the 
back end. Each spectrometer provided 16,384 channels, 
resulting in a spectral resolution of 61 kHz, equivalent to a velocity 
resolution of about 0.16~km~s$^{-1}$ for \twCO\ and 0.17~km~s$^{-1}$ 
for \thCO\ and C$^{18}$O. The half-power beam width (HPBW) of the telescope 
was about $50''$, and the pointing accuracy of the telescope was greater 
than $4''$ in the observing epoch.
We used the on-the-fly observing technique
to map the 0\fdg9$\times$0\fdg9 area centered at 
($l$=22\fdg75, $b$=$-$0\fdg25)
with a scan speed of 50$''$~s$^{-1}$ and a step of 15$''$ along the 
Galactic longitude and latitude. 
The mean rms noise level of the brightness
temperature ($T_{\rm R}$) was about 0.5~K for \twCO\ and
0.3~K for \thCO\ and C$^{18}$O. All of the CO data used in this
study are expressed in brightness temperature. Here we adopt
the main beam efficiency $\eta_{\rm mb}=$ 0.44 for
\twCO\ and 0.48 for \thCO\ and C$^{18}$O
($T_{\rm R}=T_{\rm A}/(f_{\rm b}\times\eta_{\rm mb})$,
assuming a beam filling factor of $f_{\rm b}\sim$~1).

\begin{figure}
\includegraphics[trim=10mm 0mm 0mm 90mm,scale=0.45,angle=0]{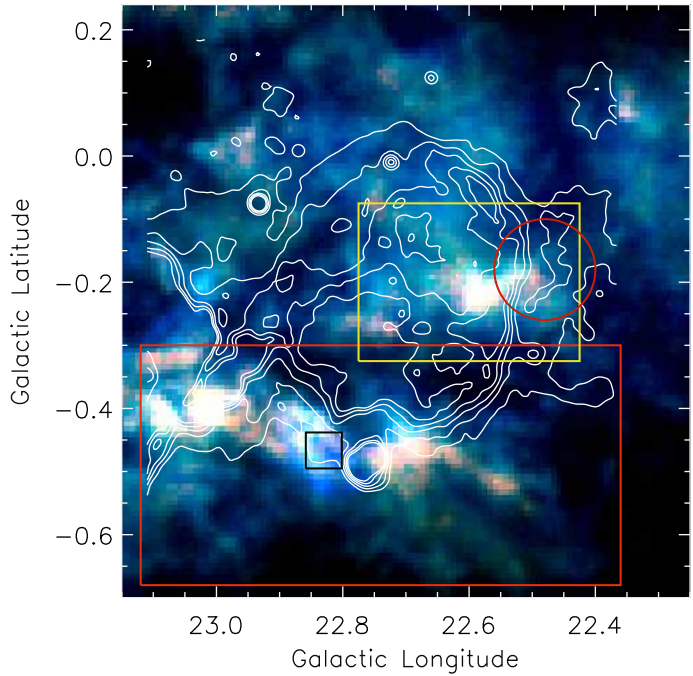}
\caption{\twCO\ ($J$=1--0; blue), \thCO\ ($J$=1--0; green), and
C$^{18}$O ($J$=1--0; red) intensity maps in the 71--86~km~s$^{-1}$
interval with a linear scale of SNR G22.7$-$0.2
overlaid with the VGPS 1.4~GHz radio continuum emission contours.
The black box indicates the region from which the spectra of the shocked
gas are extracted (see Figure~3). The red rectangle and the yellow
rectangle show the region of GMC G23.0$-$0.4W and GMC G22.6$-$0.2,
respectively. The red circle shows the location of HESS J1832--093
\citep{2013arXiv1308.0475L}. SNR W41 (G23.3$-$0.3) is located to
the left of SNR G22.7$-$0.2.
\label{fig1}}
\end{figure}

We also used the \twCO~($J$=2--1) line (at 230.538~GHz), which
was observed during 2010 January and February using the 3 m submillimeter 
telescope at the K\"{o}lner Observatory for Submillimeter Astronomy (KOSMA) 
in Switzerland. An SIS receiver and a medium-resolution 
acoustic--optical spectrometer (AOS) spectrometer were used.
The map of 0.33 deg$^2$ was centered at $(\RA{18}{33}{19},\Dec{-09}{10}{42})$ 
with a grid spacing of 1$'$. The HPBW of the telescope was 130$''$ and 
the main beam efficiency was about 0.54 during our observation.
The AOS bandwidth and velocity
resolution were about 300 MHz and 0.2~km~s$^{-1}$, respectively.

All of the CO data were reduced using the GILDAS/CLASS
package developed by IRAM\footnote{http://www.iram.fr/IRAMFR/GILDAS}.
Finally, the baseline-corrected spectra of CO ($J$=1--0) were converted to
three-dimensional cube data with a grid spacing of $30''$ and a velocity 
channel separation of 0.5~km~s$^{-1}$ for subsequent analysis. The
spectra of CO ($J$=2--1) were converted to $1'\times1'\times0.5$~km~s$^{-1}$
cube data.

The VLA Galactic Plane Survey \citep[VGPS,][]{2006AJ....132.1158S} radio 
continuum emissions were also used for comparison.

\section{RESULTS}
\subsection{MC Distribution Toward SNR G22.7$-$0.2}
In the FOV of SNR G22.7$-$0.2, the \twCO~($J$=1--0)
is in a broad velocity range between 0 and 130~km~s$^{-1}$,
characterized by multiple peaks because the remnant is near
the inner Galaxy. After checking the CO emission
intensity maps channel by channel,  
we found a morphological correlation between the 71--86~km~s$^{-1}$
MCs and SNR G22.7$-$0.2. 

\begin{figure}
\includegraphics[trim=0mm 0mm 0mm 125mm,scale=0.5,angle=0]{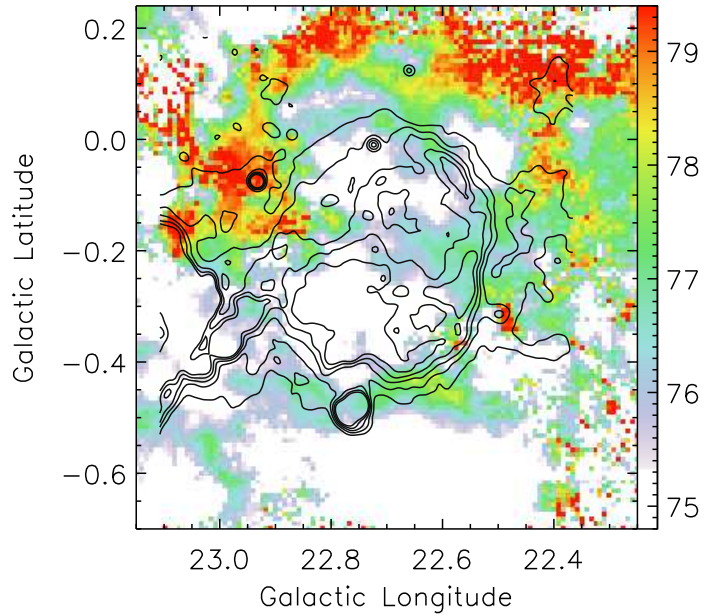}
\caption{Intensity-weighted mean velocity (first moment) map of
the \thCO\ ($J$=1--0) emission on the interval of 70--81~km~s$^{-1}$
of SNR G22.7$-$0.2 overlaid with the VGPS 1.4~GHz
radio continuum emission contours. The color scheme is adjusted to
highlight the velocity range of 75--79~km~s$^{-1}$.
\label{fig2}}
\end{figure}

The 71--86~km~s$^{-1}$ MCs were also called the molecular cloud
complex [23,78] based on 1.2m CO survey \citep{1986ApJ...305..892D}.
On the other hand, \cite{2014A&A...569A..20M} identified this GMC
as G23.3$-$0.3 at a distance of 4--5~kpc \citep{2006ApJ...643L..53A}.
Based on our new CO observations (see Section 2), we identified the
molecular cloud complex as GMC G23.0$-$0.4 ($\VLSR\sim$77~km~s$^{-1}$).
The GMC G23.0$-$0.4 centered at ($l$=23\fdg0,$b$=$-0$\fdg4) covers about 
1\fdg1$\times$0\fdg4 area in \thCO\ emission and it 
generally shows the filamentary structure.

\begin{deluxetable*}{cccccccc}
\tabletypesize{\scriptsize}
\tablecaption{Properties of GMCS }
\tablehead{
\colhead{\begin{tabular}{c}
GMC \\
     \\
\end{tabular}} &
\colhead{\begin{tabular}{c}
$T_{\rm ex}$(CO) \\
(K)              \\
\end{tabular}} &
\colhead{\begin{tabular}{c}
$\tau$(\thCO) \\
              \\
\end{tabular}} &
\colhead{\begin{tabular}{c}
Size          \\
($'\times'$)  \\
\end{tabular}} &
\colhead{\begin{tabular}{c}
$\VLSR$(peak) \\
(km~s$^{-1}$) \\
\end{tabular}} &
\colhead{\begin{tabular}{c}
Column Density$^{\mathrm {a}}$\\
(10$^{22}$cm$^{-2}$)\\
\end{tabular}} &
\colhead{\begin{tabular}{c}
Mass$^{\mathrm {a,b}}$     \\
($10^{5}\Msun$)\\
\end{tabular}} &
\colhead{\begin{tabular}{c}
Density$^{\mathrm {a,b,c}}$      \\
(cm$^{-3}$)\\
\end{tabular}}
}
\startdata
G22.6$-$0.2 & 18 & 0.27 & 14$\times$10 & 76.4 & 2.7/2.3 & 1.4$\du^2$/1.2$\du^2$ & 680$\du^{-1}$/580$\du^{-1}$ \\
G23.0$-$0.4W& 23 & 0.12 & 40$\times$22 & 76.6 & 2.6/1.9 & 8.2$\du^2$/5.7$\du^2$ & 330$\du^{-1}$/240$\du^{-1}$
\enddata
\tablecomments{
$^{a}$ See text for the two methods (LTE/X-factor) used for the calculation.
$^{b}$ Parameter $\du$ is the distance to the cloud in units of 4.4~kpc.
$^{c}$ Assuming a depth of 10$'$ for GMC G22.6$-$0.2 and 20$'$ for GMC
G23.0$-$0.4W along the line of sight.}
\end{deluxetable*}

Figure 1 shows a portion of GMC G23.0$-$0.4 that covers the SNR G22.7$-$0.2.
The composite image displays the intensity map of the CO lines integrated
in the 71--86~km~s$^{-1}$ interval (\twCO\ $J$=1--0 in blue, \thCO\ $J$=1--0
in green, and C$^{18}$O $J$=1--0 in red).
The brightest portions of the three lines are found in the region of
the west part of the GMC G23.0$-$0.4 (about 1\fdg1$\times$0\fdg4 in size) and
GMC G22.6$-$0.2 (about 0\fdg3$\times$0\fdg2 in size).
Hereafter, we regard the west part of GMC G23.0$-$0.4 as G23.0$-$0.4W.
In GMC G23.0$-$0.4W (Figure~1), the CO emission shows a filamentary structure
that nicely surrounds and folds the southern region of SNR G22.7$-$0.2.
The GMC G22.6$-$0.2 is located in the west of the remnant (see the yellow
rectangle in Figure~1) along with HESS J1832-093 (see the red circle in Figure~1).
Several relatively faint filamentary structures are also seen around
the northern part of the remnant.
To investigate the detailed
structures of the MCs in the vicinity of SNR G22.7$-$0.2, we made an
intensity-weighted velocity (the first moment) map of \thCO\ emission in the
velocity range of 70--81~km~s$^{-1}$ (see Figure~2). In Figure~2,
we can see the same filamentary structure in the GMC G23.0$-$0.4W.
The filament surrounds SNR G22.7$-$0.2 and has a velocity
of $\sim$76--78~km~s$^{-1}$.
In the northern border, the higher velocity
($\VLSR\sim$79~km~s$^{-1}$) of the molecular gas inherent to the SNR
could be contaminated by other clouds at 80--90~km~s$^{-1}$.
Similar high velocity gas is visible roughly in the top panel of Figure~11 of
\cite{2014A&A...569A..20M}.
The \thCO\ ($J$=1--0) velocity coded image of SNR G22.7$-$0.2 generally
seems to show a cavity structure.

\begin{figure}
\includegraphics[trim=-5mm 0mm 0mm 0mm,scale=0.3,angle=270]{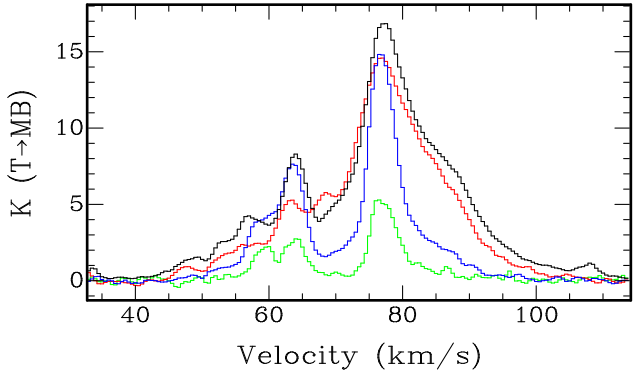}
\caption{\twCO ($J$=1--0; black), \thCO ($J$=1--0; blue, multiplied
by a factor of three),  C$^{18}$O  ($J$=1--0;  green, multiplied by a
factor of six), and  \twCO ($J$=2--1; red) spectra of the shocked gas
extracted from the region indicated in Figure~1 (black box).
\label{fig3}}
\end{figure}

\begin{figure}
\includegraphics[trim=5mm 0mm 0mm 0mm,scale=0.5,angle=0]{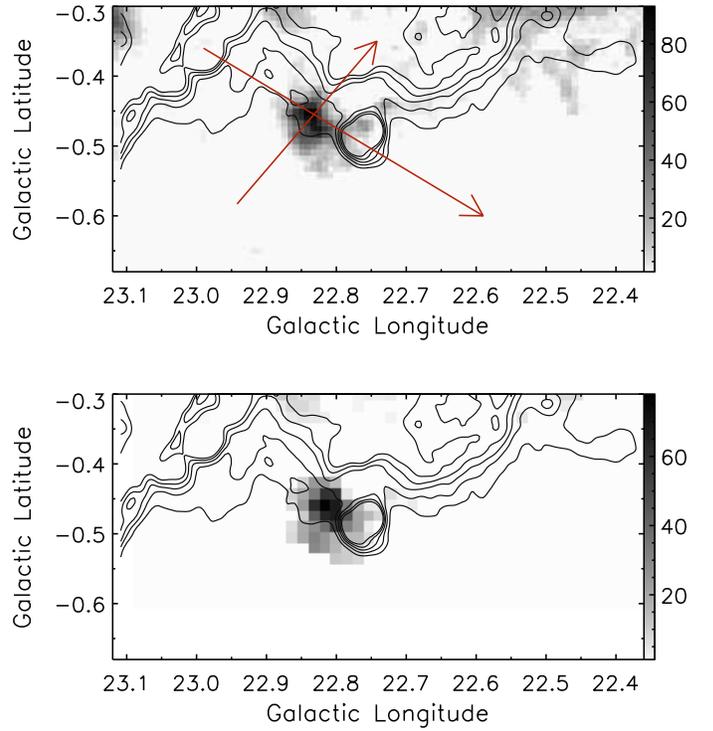}
\caption{Top: the \twCO\ ($J$=1--0) intensity map in the
86--99~km~s$^{-1}$ interval displayed the linear scale and overlaid with
the VGPS 1.4~GHz radio continuum emission contours. The red arrows show the
direction of the PV diagrams (see Figure~5).
Bottom: \twCO\ ($J$=2--1) intensity maps in the 86--99~km~s$^{-1}$
interval.\label{fig4}}
\end{figure}

We show the physical properties of the GMCs G23.0$-$0.4W and G22.6$-$0.2
in Table~1. Here two methods have been used in the
derivation of the column densities and the masses of the GMCs.
In the first method, on the assumption of local
thermodynamic equilibrium (LTE) and the \twCO\ ($J$=1--0) line being
optically thick, we can derive the excitation temperature from the peak
radiation temperature of the \twCO\ ($J$=1--0). The \thCO\ ($J$=1--0) emission
is optically thin and the \thCO\ column density is converted to the
H$_2$ column density using $N$(H$_2$)/$N(^{13}$CO) $\approx7\E{5}$
\citep{1982ApJ...262..590F}.
In the second method, the H$_2$ column density is
estimated by adopting the mean CO-to-H$_2$ mass conversion factor
$1.8\E{20}$~cm$^{-2}$K$^{-1}$km$^{-1}$s \citep{2001ApJ...547..792D}. 
In the estimate of the mass of the
GMCs, a mean molecular weight per H$_2$ molecule of
2.76 has been adopted. The distance of the GMCs
was adopted as 4.4~kpc (see Section 4.1). 

\begin{figure*}
\includegraphics[trim=-20mm 0mm 0mm 0mm,scale=0.8,angle=0]{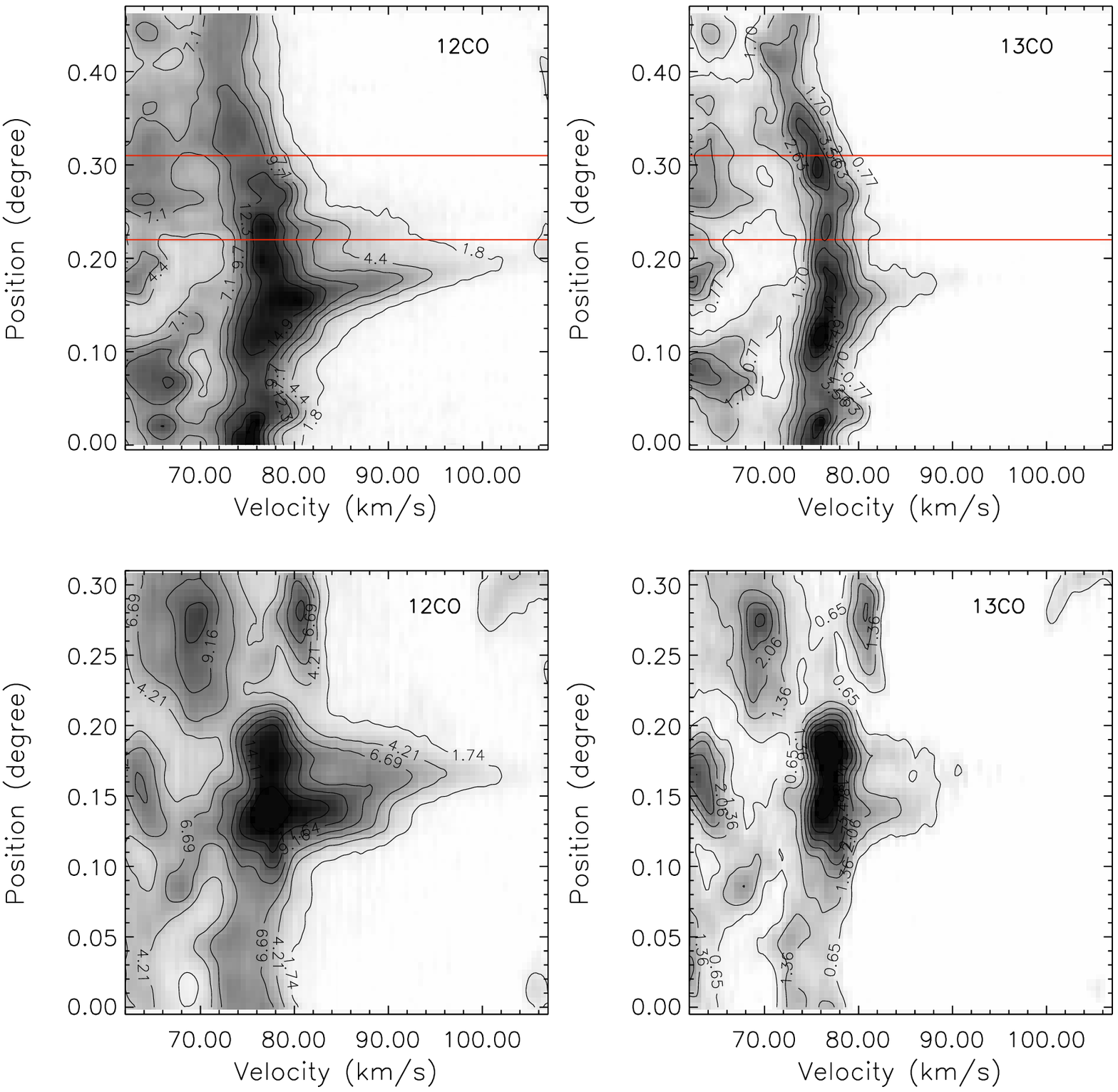}
\caption{PV diagrams of the \twCO\ ($J$=1--0) and
\thCO\ ($J$=1--0) emission toward
the shocked gas. The position is measured along the long arrow
(($l$=22\fdg990, $b$=$-$0\fdg360) to ($l$=22\fdg590, $b$=$-$0\fdg590), upper
panels) and the short arrow (($l$=22\fdg942, $b$=$-$0\fdg583) to
($l$=22\fdg742, $b$=$-$0\fdg350), lower panels)
with a width of 1\farcm5
(see Figure~4). The red lines in the upper panels mark the spatial region
of \HII\ region G22.76$-$0.485 (ID 06 in Table~3). \label{fig5}}
\end{figure*}

We found that the derived physical parameters of the LTE
method are similar to that of the X-factor method (Table~1).
On the other hand,
the mean optical depth of the \thCO\ ($J$=1--0) emission of 
GMC G23.0$-$0.4W is a half of that of GMC G22.6$-$0.2.

\subsection{Shocked Molecular Gas in the Remnant}

To search for the kinematic evidence for the interaction between SNR
G22.7$-$0.2 and the ambient MCs, we analyzed the spectra of the
\twCO\ ($J$=1--0 and $J$=2--1) emission of the remnant in detail.
The spectra of the \thCO\ ($J$=1--0) were also used for comparison,
considering that the \thCO\ line is more optically thin than the \twCO\ lines
and thus can be used to show the relatively undisturbed component.
We found the clear line broadening structure of the \twCO\ emission
in the southeastern region of the remnant, which extends from the peak velocity
of $\sim$77~km~s$^{-1}$ to $\sim$110~km~s$^{-1}$.
The redshifted profile indicates that the molecular gas of the 77~km~s$^{-1}$
MC is mainly located behind the expanding SNR.
The spectra are shown in Figure~3 and the spectral extraction region
is indicated by the black box in Figure~1.
The \twCO\ ($J$=1--0 and
$J$=2--1) line broadenings provide kinematic evidence for the
interaction between SNR G22.7$-$0.2 and the ambient molecular gas
with a system velocity of $\sim$77~km~s$^{-1}$.
Figure~4 displays the intensity maps of the shocked gas in the
86--99~km~s$^{-1}$ interval. The distribution of the shocked gas
seems to extend along
the southeastern radio border of SNR G22.7$-$0.2.

We did not find the broadening profile of the \twCO\
emission in the region of GMC G22.6$-$0.2, although it has a similar
velocity to that of the southern cloud in GMC G23.0-0.4W and it is projectively
located inside the remnant.
There may be two possible reasons to explain this.
First, the GMC G22.6$-$0.2 is not actually in physical contact with SNR
G22.7-0.2, but is instead in the foreground or background of the remnant.
Second, the SNR's shock is interacting with the molecular gas
of the GMC G22.6$-$0.2. In this case, the broadened lines are probably
contaminated by other velocity components between 70~km~s$^{-1}$
and 90~km~s$^{-1}$. This limited our investigation for the SNR--MC
interaction in CO emission. Other observations are needed to clarify
the issue.

Figure~5 shows the position--velocity (PV) diagrams of the shocked gas
along and perpendicular to the southeastern radio boundary of SNR
G22.7$-$0.2 (see the red arrows in Figure~4). We can discern
the quiescent component in the 70--81~km~s$^{-1}$ interval (see the \thCO\
PV diagrams in the right panels) and the 
shocked component in the 83--106~km~s$^{-1}$ interval (see the \twCO\
PV diagrams in the left panels). 
The highest velocity of the shocked gas as revealed by the \twCO\ 
PV diagrams is extended to $\sim$110~km~s$^{-1}$, indicative of  
$\sim$30~km~s$^{-1}$ shock velocity at least. 
The value of the broadening velocity is comparable to that
of SNR W44 \cite[$>$25~km~s$^{-1}$;][]{2004AJ....127.1098S} and 
SNR IC~443 \cite[10--100~km~s$^{-1}$;][]{2005ApJ...620..758S}.
On the other hand, the value of the broadening velocity of SNR G22.7$-$0.2 is 
larger than that found in other SNR--MC interaction systems
\cite[several kilometers per second for SNR Kes~75 and SNR 3C~396;][]
{2009ApJ...694..376S,2011ApJ...727...43S}.

We also made an intensity-weighted velocity dispersion 
(the second moment) map of the \thCO\ emission in the velocity range of 
70--81~km~s$^{-1}$ toward the molecular gas in GMC G23.0$-$0.4W (see Figure~6). 
This roughly shows the velocity dispersion in the given velocity
range of 70--81~km~s$^{-1}$: the higher the velocity dispersion,
the greater the range of velocities. Some interesting
structures are found to be located in the south of the remnant. In Figure~6,
a filamentary structure (the red arrow) and a concave structure (the
blue dotted line) seem to be associated with the remnant.
The arrow is roughly located at the inner border of the SNR
shell as seen in Figures 1 and 2.
The \twCO\ PV diagram of the filament also shows the line broadening  
structure (see the 84--90~km~s$^{-1}$ emission in the left panel of Figure~7
) compared to the \thCO\ PV diagram (right panel of Figure~7),
which is likely due to the shock--MC interaction in the region.
We will discuss the possible origin of the concave structure in Section 4.2.

\begin{figure}
\includegraphics[trim=10mm 0mm 0mm 0mm,scale=0.85,angle=0]{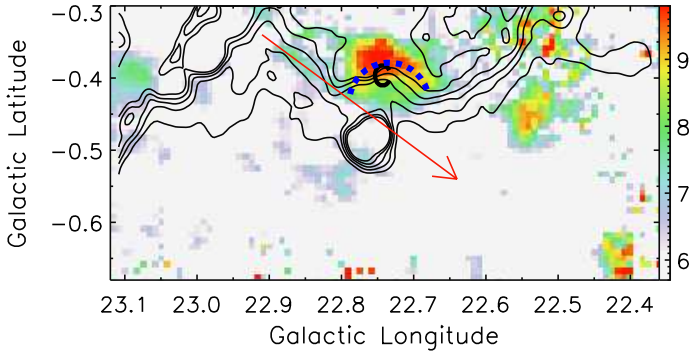}
\caption{Intensity-weighted mean velocity dispersion
(second moment) map of the \thCO\ ($J$=1--0) emission in the interval
of 70--81~km~s$^{-1}$ of the molecular gas in GMC G23.0$-$0.4W,
overlaid with the VGPS 1.4~GHz radio continuum emission contours.
The color scheme is adjusted to highlight the enhancement of the
velocity dispersion. The red arrow indicates the direction of the
PV diagrams (see Figure~7) and the blue dotted
line indicates the concave structure. The character ``C"
indicates the location of the stellar cluster GLIMPSE9
\citep{2010ApJ...708.1241M}. \label{fig6}}
\end{figure}

\begin{figure*}
\includegraphics[trim=-20mm 0mm 0mm 0mm,scale=0.7,angle=0]{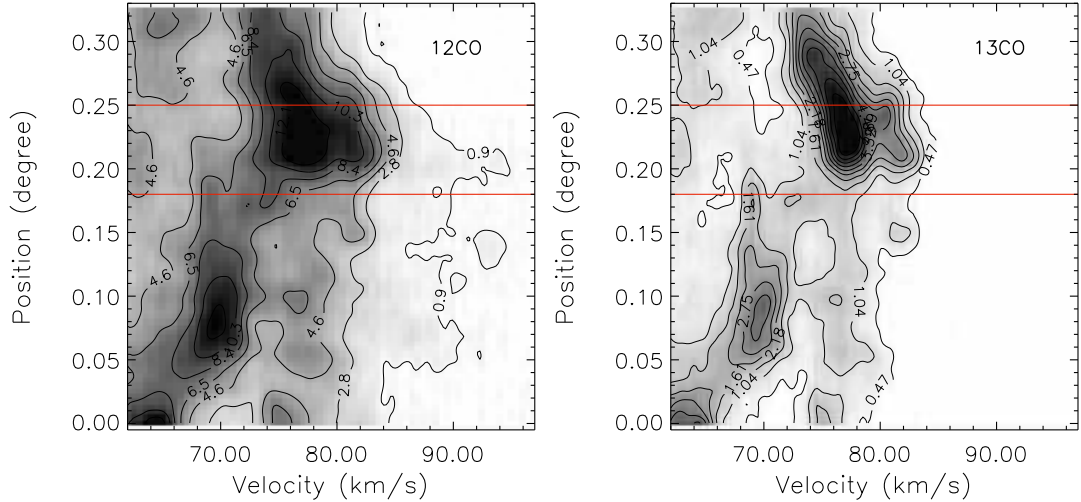}
\caption{PV diagrams of the \twCO\ ($J$=1--0) and
\thCO\ ($J$=1--0) emission along
the molecular gas slice, which is located along the southeastern bright
radio shell of the remnant. The position is measured along the arrow
(($l$=22\fdg910, $b$=$-$0\fdg340) to ($l$=22\fdg640, $b$=$-$0\fdg520), see
Figure~6) with a width of 1\farcm5. The red lines mark the spatial region
of \HII\ region G22.76$-$0.485 (ID 06 in Table~3). \label{fig7}}
\end{figure*}

The CO component in the velocity of $\sim$60--70~km~s$^{-1}$ limited
the investigation for the blueshifted wing of the shocked gas.
However, the enhancement of the \twCO\ ($J$=2--1) in the interval
of 68--74~km~s$^{-1}$ probably indicates the shock--MC interaction
(see Figure~3). More observations are needed to clarify the characteristics
of the shocked gas in the region.

Weak \thCO~($J$=1--0) emission has been detected from the
high-velocity wings in the region of the shocked gas (see Figures~3 and 5),
which probably indicates that the \twCO~($J$=1--0) line wings
are not optically thin. Coincidentally, SNR IC~443, another SNR--MC system,
also displays very weak wings of the \thCO~($J$=1--0) line in the velocity interval
of $-50$ to $-10$~km~s$^{-1}$ toward the shocked gas of the clump B
\citep[see Figure~15 in][]{2010SCPMA..53.1357Z}. It shows that the
\twCO\ emission of the broadening for shocked gas probably has a
median optical depth for some cases.

If the excitation temperatures and the beam filling factors of
the \twCO~($J$=1--0) and \thCO~($J$=1--0) lines are equal, the average optical
depth of the shocked gas can be estimated from the ratio of the
integrated intensity over the high-velocity emission for \twCO/\thCO\
\citep{1984ApJ...284..176S}.
The integrated intensity ratio value is about 20.9 for a $5'\times5'$
shocked region centered at ($l$=22\fdg833,$b$=$-$0\fdg467).
Assuming the optical depth of \thCO~($J$=1--0)
emission $\tau(13)\ll$1 in the line wing and the isotopic ratio
$^{12}$C/$^{13}$C= $\tau(12)$/$\tau(13)$=41
\citep{2005ApJ...634.1126M},
we get the optical depth of $\tau(12)\sim$1.6 for the shocked
high velocity \twCO~($J$=1--0) gas.
We note that the optical depth estimated for the shocked gas
is close to that of the stellar cluster GLIMPSE9 in $K_{\rm S}$
band \citep{2010ApJ...708.1241M}.
The column density of the shocked gas in the interval of 83--106~km~s$^{-1}$
(Figure~5) can be estimated as 2.4$\times$10$^{21}$~cm$^{-2}$.
The mean density of the shocked gas is about 120$\du^{-1}$~cm$^{-3}$
assuming the depth of 5$'$ along the line of sight,
where $\du=d/(4.4$~kpc) is the distance scaled with 4.4~kpc (see Section 4.1).
In the above calculation, a uniform excitation temperature ($T_{\rm ex}$=23~K,
Table~1) and a conversion factor $N$(H$_2$)/$N(^{12}$CO) $\approx1\E{4}$
\citep{1989ApJS...69..271H} are assumed. Therefore, the mass,
momentum, and energy of the high velocity shocked gas
are estimated to be 2.2$\du^2\times$10$^{3}\Msun$,
3.3$\du^2\times$10$^{4}\Msun$~km~s$^{-1}$,
and 5.0$\du^2\times$10$^{48}$~erg, respectively.
These values should be regarded as
lower limits since it is very likely that the excitation temperature is
$T_{\rm ex}$(CO wing)$\geq$23~K at the shock-heated region and that the
optical depth of \twCO~($J$=1--0) emission is $\tau(12)\geq$1.6 for the
region of the high column density.
Additionally, the low-velocity component of the shocked gas is not
accounted for because of the contamination of the line center emission.

\section{DISCUSSION}
\subsection{Association of SNR G22.7$-$0.2 with the 
77~km~s$^{-1}$ MCs and the Distance to the Remnant}
In Section 3, we showed spatial and kinematic evidence
for the association between SNR G22.7$-$0.2 and the 
$\VLSR\sim$77~km~s$^{-1}$ MCs. First, the molecular gas in GMC 
G23.0$-$0.4W at a radial velocity of 71--86~km~s$^{-1}$
is found to surround the southern radio shell of the remnant 
(Figures~1 and 2). The map of the \thCO\ velocity 
distribution seems to display a cavity structure in which
SNR G22.7$-$0.2 is embedded (Figure~2). Second, the 83--110~km~s$^{-1}$
red wing of the \twCO\ ($J$=1--0 and $J$=2--1) emission
is found in a region of GMC G23.0$-$0.4W ($l$=22\fdg833, $b$=$-$0\fdg467),
providing convincing kinematic evidence for the SNR--MC interaction
(Figures~3 and 5). The \twCO\ PV diagram of a filament along the 
southeastern radio shell (($l$=22\fdg850, $b$=$-$0\fdg384),
see the red arrow in Figure~6) also 
displays the broadening structure, which further supports
the interaction between the remnant and the 77~km~s$^{-1}$ MCs
(the left panel in Figure~7).

\begin{deluxetable}{ccccc}
\tabletypesize{\scriptsize}
\tablecaption{Dust-Continuum-Identified MC Clumps toward SNR G22.7$-$0.2 }
\tablehead{
\colhead{\begin{tabular}{c}
ID \\
   \\
\end{tabular}} &
\colhead{\begin{tabular}{c}
$l$    \\
(deg)  \\
\end{tabular}} &
\colhead{\begin{tabular}{c}
$b$    \\
(deg)  \\
\end{tabular}} &
\colhead{\begin{tabular}{c}
$\VLSR$$^{\mathrm {a}}$      \\
(km~s$^{-1}$)\\
\end{tabular}} &
\colhead{\begin{tabular}{c}
Distance$^{\mathrm {a}}$      \\
(kpc)           \\
\end{tabular}}
}
\startdata
a  &   22.474 & $-$0.223 &   75.9  &  $4.70^{+0.36}_{-0.36}$  \\
b  &   22.494 & $-$0.207 &   75.8  &  $4.58^{+0.26}_{-0.28}$  \\
c  &   22.504 & $-$0.197 &   76.8  &  $4.64^{+0.28}_{-0.32}$  \\
d  &   22.534 & $-$0.193 &   75.6  &  $4.66^{+0.32}_{-0.34}$  \\
e  &   22.694 & $-$0.454 &   77.9  &  $4.70^{+0.26}_{-0.30}$  \\
f  &   22.836 & $-$0.418 &   73.4  &  $4.50^{+0.30}_{-0.34}$  \\
g  &   22.870 & $-$0.408 &   75.4  &  $4.54^{+0.34}_{-0.34}$  \\
h  &   22.878 & $-$0.434 &   76.8  &  $4.44^{+0.30}_{-0.30}$  \\
i  &   22.906 & $-$0.466 &   76.8  &  $4.60^{+0.32}_{-0.34}$
\enddata
\tablecomments{
$^{a}$ See Section 3 and Table~3 in \cite{2013ApJ...770...39E}.
}
\end{deluxetable}

\begin{deluxetable*}{ccccccc}
\tabletypesize{\scriptsize}
\tablecaption{Overlapping \HII\ regions toward SNR G22.7$-$0.2 }
\tablehead{
\colhead{\begin{tabular}{c}
ID \\
   \\
\end{tabular}} &
\colhead{\begin{tabular}{c}
Name \\
     \\
\end{tabular}} &
\colhead{\begin{tabular}{c}
$l$    \\
(deg)  \\
\end{tabular}} &
\colhead{\begin{tabular}{c}
$b$    \\
(deg)  \\
\end{tabular}} &
\colhead{\begin{tabular}{c}
Radius \\
(arcmin)\\
\end{tabular}} &
\colhead{\begin{tabular}{c}
$\VLSR$      \\
(km~s$^{-1}$)\\
\end{tabular}} &
\colhead{\begin{tabular}{c}
Ref.$^{\mathrm {a}}$\\
          \\
\end{tabular}}
	 }
\startdata
01  &  G022.398$+$0.083  &  22.398   &  $+$0.083  & \nodata &  87.8        & 3  \\
02  &  G022.724$-$0.010  &  22.724   &  $-$0.010  &  0.7    &  38.9        & 1,2 \\
03  &  G022.730$-$0.239  &  22.730   &  $-$0.239  &  1.0    &  71.1/113.7  & 1,2 \\
04  &  G022.739$-$0.303  &  22.739   &  $-$0.303  & \nodata &  69.3/112.1  & 1   \\
05  &  G022.755$-$0.246  &  22.755   &  $-$0.246  &  1.2    &  70.2/106.7  & 1,2 \\
06$^{\mathrm {b}}$  &  G022.760$-$0.485  &  22.760   &  $-$0.485  &  2.6    &  74.8        & 2,3,5 \\
07  &  G022.780$-$0.383  &  22.780   &  $-$0.383  &  1.7    &  70.0        & 1,2 \\
08  &  G022.935$-$0.072  &  22.935   &  $-$0.072  & \nodata &  71.2        & 4   \\
09  &  G022.947$-$0.315  &  22.947   &  $-$0.315  &  0.7    &  70.9        & 2,3 \\
10$^{\mathrm {c}}$  &  G022.982$-$0.356  &  22.982   &  $-$0.356  &  2.5    &  74.1        & 2,3 \\
11  &  G022.986$-$0.149  &  22.986   &  $-$0.149  &  1.9    &  76.6/100.2  & 1,2 \\
12  &  G023.029$-$0.405  &  23.029   &  $-$0.405  &  2.5    &  76.0        & 1   \\
13  &  G023.067$-$0.367  &  23.067   &  $-$0.367  & \nodata &  82.7        & 3   \\
14  &  G023.072$-$0.248  &  23.072   &  $-$0.248  & \nodata &  89.6        & 3
\enddata
\tablecomments{
$^{a}$ (1)~\citealp{2011ApJS..194...32A}; (2)~\citealp{2014ApJS..212....1A};
(3)~\citealp{1989ApJS...71..469L}; (4)~\citealp{1996ApJ...472..173L};
(5)~\citealp{2006A&A...453.1003T}.
$^{b}$ The source is a SNR candidate (G022.7583$-$0.4917) based on the
Multi-Array Galactic Plane Imaging Survey \citep{2006AJ....131.2525H}.
$^{c}$ The source is a SNR candidate (G022.9917$-$0.3583) based on the
Multi-Array Galactic Plane Imaging Survey \citep{2006AJ....131.2525H}.
}
\end{deluxetable*}

The association of SNR G22.7$-$0.2 with the MCs at the
systemic velocity, $\sim$77~km~s$^{-1}$, enables us to place the
SNR at a near kinematic distance of 4.4$\pm$0.4~kpc by using 
the Galactic rotation curve model of
\cite{2014ApJ...783..130R}. We also exclude the far distance of
the 77~km~s$^{-1}$ MCs based on the \mbox{H\,\textsc{i}} 
self-absorption method \citep{2009ApJ...699.1153R}. The 4.4$\pm$0.4~kpc 
kinematic distance with a system velocity of $\sim$77~km~s$^{-1}$, roughly 
places the remnant at the near side of the Scutum--Crux arm 
\citep[see the distance of the Galactic arm in][]{1993ApJ...411..674T,2004ApJS..154..553S}.
It is also consistent with the distance of the dense clumps in
the GMC with velocity of $\sim$77~km~s$^{-1}$ (Table~2),
which is estimated for dust-continuum-identified MC clumps from the
Bolocam Galactic Plane Survey in the inner Galaxy \citep{2013ApJ...770...39E}.
These MC clumps are probably the cradle of massive star formation
in the GMC with a velocity of $\sim$77~km~s$^{-1}$. 
We note that the distance of SNR G22.7$-$0.2 is close to
that of the nearby SNR W41 (see the radio contours in the 
east of Figure~1). \cite{2008AJ....135..167L} estimated the 
kinematic distance of 3.9--4.5~kpc for SNR W41 based on the 
\mbox{H\,\textsc{i}} and \thCO\ data. 
Recently, \cite{2013ApJ...773L..19F} report on the discovery
of 1720~MHz OH line emission of the radio continuum
of SNR W41 with the radial velocity of 74~km~s$^{-1}$, which is
strong evidence for a SNR--MC interaction system.
The radial velocity of the OH emission coincides with the system
velocity of the east part of GMC G23.0$-$0.4, indicative of
the interaction between SNR W41 and the GMC. Moreover,
the east part of GMC G23.0$-$0.4 with $\VLSR\sim$77~km~s$^{-1}$ extends
toward SNR W41 and it covers the projected area of the remnant. The 
east part of GMC G23.0$-$0.4 is probably partly responsible for the 
bright VHE emission of HESS J1834$-$087 (see Section 4.3). 
Since SNRs G22.7$-0.2$ and W41 are both interacting with the 
GMC G23.0$-$0.4, the two SNRs are at a
similar distance.

The radius of SNR G22.7$-0.2$ is about 18$\du$~pc and
no extended X-ray emission has been detected up to now. Assuming the 
$\sim$100~km~s$^{-1}$ expansion velocity of the shock in the
radiative phase, the age
of SNR G22.7$-$0.2 is about several tens of thousands of years.
Adopting the SNR's radius of $\sim$14$'$, the solid angle
of the shocked gas is about 0.1~sr in the southeastern boundary of 
the remnant. Thus we estimated the kinetic energy
of the remnant to be at least $\sim$6.3$\du^2\times$10$^{50}$~erg,
which is consistent with the typical kinetic energy release of 
10$^{51}$~erg for a SNR.

\subsection{Relationship Between the SNR and the Overlapping 
Star-forming Regions}
There are multiple \HII\ regions in the FOV of SNR
G22.7$-$0.2. We list the physical parameters of these \HII\ regions in Table~3. 
Most of the \HII\ regions seem to be associated with the 
70--81~km~s$^{-1}$ MCs, e.g., GMC G23.0$-$0.4W (see Figure~8). The
\HII\ region G022.760$-$0.485 with a systemic velocity of 74.8~km~s$^{-1}$
(ID 6 in Table~3) is located in the south of SNR G22.7$-$0.2,
in which the radio emission of the remnant shows a concave structure. 
A spectrophotometric distance of 4.2$\pm$0.4~kpc was derived for the cluster
GLIMPSE9 ($l$=22\fdg756, $b$=$-$0\fdg400) \citep{2010ApJ...708.1241M} and 
of $\sim$4.6~kpc for molecular complex \citep{2014A&A...569A..20M}. 
The cluster is located in the concave structure (see the character ``C" 
in Figure~6).
\cite{2009ApJ...693..424B} measured the distance of 
$4.59^{+0.38}_{-0.33}$~kpc based on
the trigonometric parallax for the massive star-forming region
G23.01$-$0.41, which is near the \HII\ region G023.029$-$0.405
(ID 12 in Table~3, $\VLSR\sim$76~km~s$^{-1}$). 
\cite{2013ApJ...770...39E} presented a new distance estimation for
dust-continuum-identified MC clumps, nine of which are probably associated
with the 77~km~s$^{-1}$ GMCs (Table~2, Figure~8). 
Therefore, these objects are likely associated with the GMC with velocity
$\sim$77~km~s$^{-1}$ because of the systematic velocity coincidence.

\begin{figure}
\includegraphics[trim=10mm 0mm 0mm 125mm,scale=0.5,angle=0]{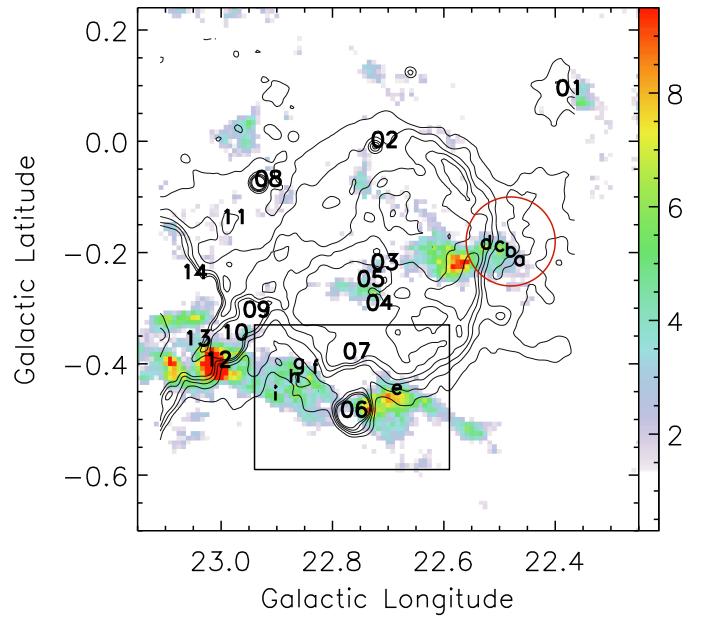}
\caption{Location of the \HII\ regions
of SNR G22.7$-$0.2. The intensity map is the
C$^{18}$O ($J$=1--0) emission in the interval of 70--81~km~s$^{-1}$,
overlaid with the VGPS 1.4~GHz radio continuum emission contours.
The characters a--i indicate the dust-continuum-identified MC clumps
in Table~2 and
ID 01--14 indicate the \HII\ regions listed in Table~3.
The red circle shows the location of HESS J1832--093 \citep{2013arXiv1308.0475L}.
The black box indicates the region of the shocked gas (see Figure~9).
\label{fig8}}
\end{figure}

As mentioned in Sections 3.2 and 4.1, we show that SNR G22.7$-$0.2 is 
interacting with the southern molecular gas in GMC G23.0$-$0.4W. The 
velocity coincidence 
of the \HII\ region G022.760$-$0.485 and the ambient 77~km~s$^{-1}$ GMC 
indicates that the \HII\ region is probably physically associated with 
SNR G22.7$-$0.2. The overlapping of the velocity 
dispersion of MCs and the radio emission in the concave 
structure (blue dotted line in Figure~6) strengthen above suggestion. 
The enhancement of the velocity dispersion in the concave structure
(Figure~6) is probably
due to the shock perturbation by SNR G22.7$-$0.2 and the \HII\ region
G022.760$-$0.485.
Moreover, the interstellar extinction and spectral type of stars in the 
core of G22.760$-$0.485 and in the proximity of the southern border of SNR 
G22.7$-$0.2 supports the association of SNR G22.7$-$0.2 and G022.760$-$0.485 
with the same GMC \citep{2014A&A...569A..20M}.
We note that a few of these radio-continuum sources may be composite sources
(e.g., \HII\ region G022.760$-$0.485).
\cite{2006AJ....131.2525H} identified G022.7583$-$0.4917 as a SNR
candidate with diameter of 5$'$ and \cite{2010ApJ...708.1241M}
measured a radio spectral index of $-0.98$ of the source (see
Region 3 in their Table~3).

\begin{figure}
\includegraphics[trim=15mm 0mm 0mm 125mm,scale=0.45,angle=0]{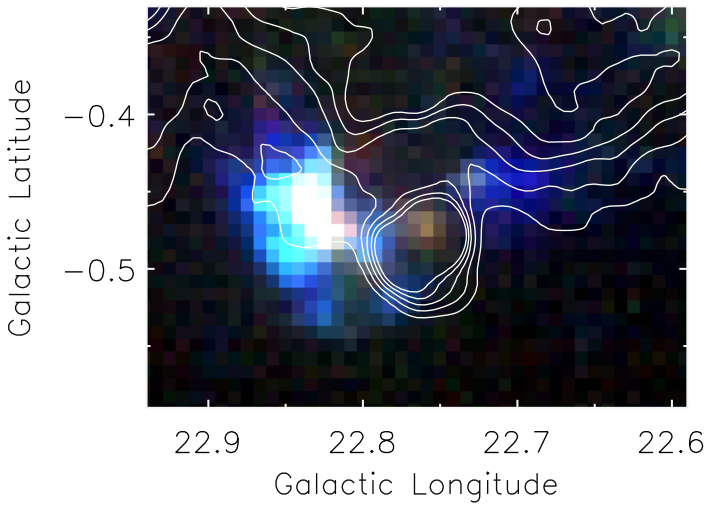}
\caption{
\twCO\ ($J$=1--0) 3-color intensity map (84--88~km~s$^{-1}$ in blue,
88--92~km~s$^{-1}$ in green, and 92--96~km~s$^{-1}$ in red) of
SNR G22.7$-$0.2, which
shows the distribution of the shocked gas in the south of the remnant.
The map shows the southern border of SNR G22.7$-$0.2 and \HII\ region
G22.76$-$0.485.
\label{fig9}}
\end{figure}

Figure~9 shows a map intensity of the shocked gas in the south of 
the remnant. We find that the shocked gas is mainly located on the eastern
edge of \HII\ region G022.760$-$0.485. A part of the faint emission of the shocked 
gas also can be discerned in the northwestern boundary of the \HII\ region.
Nevertheless, there is little emission of the shocked gas in the region 
between the remnant and the \HII\ region. 
A possible explanation is that the molecular gas in the north of \HII\
region G022.760$-$0.485 were swept away by the SNR's shock and were exhausted 
by the interaction between the SNR and the \HII\ region. 

The age of the stellar cluster, which is located in the concave 
structure (see the character ``C" in Figure~6), is about 27 to several 
Myrs \citep{2010ApJ...708.1241M}. 
On the other hand, the age of \HII\ region G022.760$-$0.485, which is 
about 5$'$ away from the stellar cluster,
is about 0.1 to several Myrs \citep{1980pim..book.....D}.
Meanwhile, \cite{2014A&A...569A..20M} estimated the age of the O-types stars
associated with GMC G23.0$-$0.4 (the GMC G23.3$-$0.3 in their 
paper) of a likely age of 5--8~Myr. Interestingly, one of star 
(ID 25 in their paper) is exactly located on the center of the \HII\ region
G022.760$-$0.485.
Presuming that the stellar cluster and \HII\ region G022.760$-$0.485
were birthed in the same MCs in the southern edge of the remnant, 
the progenitor's activities of SNR G22.7$-$0.2 seem a plausible triggering
mechanism for the new generation of star formation in the ambient GMC.

\subsection{The Very High-Energy $\gamma$-ray Source {\rm HESS}
J1832$-$093 Adjacent to the Remnant}
The MCs that SNRs are associated
or interacting with are a good probe for the hadronic process
\citep[see the recent view in][]{2014IAUS..296..170C}.
Several SNRs are detected with bright
high-energy emission by their interaction with nearby MCs, e.g., SNRs 
W28 \citep{2008A&A...481..401A} and 
IC~443 \citep{2010ApJ...712..459A,2013Sci...339..807A,
2014ApJ...788..122S}. At the west boundary of SNR G22.7$-$0.2,
HESS J1832$-$093 is detected with a significance of 5.6$\sigma$
\citep{2013arXiv1308.0475L}. 
\cite{2013arXiv1308.0475L} found an infrared source, 2MASS J18324516$-$0921545, 
around the position of a point-like X-ray source, XMMU J183245$-$0921539.
They discussed the possible origination of the VHE source.
The VHE emission is spatially 
coincident with the 1$\times10^5\Msun$ GMC G22.6$-$0.2 (Figures~1 and 8).
A possible origin of the $\gamma$-rays is hadronic interactions of the
cosmic rays with the GMC.

There is no VHE emission detected in the south of SNR
G22.7$-$0.2, where massive molecular gas in GMC G23.0$-$0.4W is located.
However, \cite{2013ApJ...773L..19F} argued that the 
correlation between the VHE $\gamma$-ray source
HESS J1834$-$087 and the 1720~MHz OH masers of
SNR W41 favors a hadronic interpretation for the VHE emission.
The 74~km~s$^{-1}$ radial velocity of the OH masers is consistent
with the velocity of GMC G23.0$-$0.4, indicating that SNR W41,
behind the GMC, is in physical contact with it \citep{2013ApJ...773L..19F}. 
\cite{2014arXiv1407.0862T} discussed the origin of the
high-energy emission of SNR W41 in detail and they also favored
the hadronic scenario from GeV to TeV extended emission.
Why is there no detectable high-energy emission in the south of
SNR G22.7$-$0.2? 
Further searches for detecting and characterizing high-energy emission in the 
vicinity of SNR G22.7$-$0.2 are needed to be performed.

\section{SUMMARY}
Millimeter and submillimeter CO studies have been performed 
on SNR G22.7$-$0.2.
We conclude the main results of our analysis as follows.

1. We have found spatial and kinematic evidence to support the association
between SNR G22.7$-$0.2 and the nearby 77~km~s$^{-1}$ MCs. The intensity map
of the molecular gas in GMC G23.0$-$0.4W in 75--79~km~s$^{-1}$ displays  
filamentary structures surrounding the southern boundary of the remnant.
The molecular gas ($l$=22\fdg833,$b$=$-$0\fdg467) shows a 
redshifted broadening (77--110~km~s$^{-1}$) in
the \twCO\ ($J$=1--0 and $J$=2--1) line emission, which indicates
convincing kinematic evidence for SNR--MC interaction.

2. We place SNR G22.7$-$0.2 at a kinematic distance of 4.4$\pm$0.4~kpc 
based on the association between the remnant and the 77~km~s$^{-1}$
GMC G23.0$-$0.4. The SNR is located at the near side of the 
Scutum-Crux arm, where several \HII\ regions 
are evolving in the complicated MC environment.

3. We suggest that the overlapping \HII\ region G022.760$-$0.485 with
a radial velocity $\sim$74.8~km~s$^{-1}$ is possibly associated
with SNR G22.7$-$0.2 and is likely to be triggered by the stellar 
winds from the massive progenitor of the remnant.

4. SNRs G22.7$-$0.2 and W41 are both interacting with 
$\VLSR\sim$77~km~s$^{-1}$ GMC G23.0$-$0.4, indicating that they
very likely have a similar distance.

\acknowledgments
The authors acknowledge the staff members of the Qinghai Radio
Observing Station at Delingha for their support of the observations.
We thank the anonymous referee for valuable advice and comments.
This work is supported by NSFC grants 11103082, 11233001, 11233007,
and 10725312. Y.S. acknowledges support from grant BK2011889. 
The work is a part of the Multi-Line Galactic Plane Survey in CO and its
Isotopic Transitions, also called the Milky Way Imaging Scroll
Painting, which is supported by the Strategic Priority Research Program,
the Emergence of Cosmological Structures of the Chinese Academy of 
Sciences, grant No. XDB09000000.

\bibliographystyle{apj}
\bibliography{references}

\end{document}